\documentclass[submission,copyright,creativecommons]{eptcs}
\providecommand{\event}{LFMTP 2015} 

\newcommand{\hideshow}[1]{{\mbox{}}}
\newcommand{\dotminus}{\mathbin{\scriptstyle\dot{\smash{\textstyle-}}}}

\usepackage{multicol}
\usepackage{proof}
\usepackage{stmaryrd}
\usepackage{latexsym}

\usepackage{amsmath}
\usepackage{amssymb}
\usepackage{amsfonts}
\usepackage{amsthm}

\usepackage{esvect}
\usepackage{alltt}
\usepackage{bussproofs}
\usepackage[all]{xy}
\usepackage{tabularx}

\usepackage{multicol}
\usepackage{color}
\usepackage[usenames,dvipsnames]{xcolor}

\theoremstyle{plain}
\newtheorem{theorem}{Theorem}[section]
\newtheorem{lemma}[theorem]{Lemma}

\theoremstyle{definition}
\newtheorem{definition}{Definition}[section]

\newcommand{\E}[2]{\ensuremath{{\epsilon}}}

\newcommand{\LF}  {\mbox{$\mathsf {LF}$}}

\newcommand{\LFP}   {\mbox{$\mathsf{LF}_{\mathcal P}$}}

\newcommand{\LLFP}  {\mbox{$\mathsf{LLF}_{\mathcal P}$}}

\newcommand{\CLLFP} {\mbox{$\mathsf{CLLF}_{\mathcal P}$}}
\newcommand{\CLLFPQ} {\mbox{$\mathsf{CLLF}_{\mathcal P?}$}}
\newcommand{\FPST}   {\mbox{$\mathsf{FPST}$}}

\newcommand {\ala}       {{\textit{\`a la}}}

\newcommand {\eg}        {{\textit{e}.\textit{g}.}}

\newcommand {\ie}        {{\textit{i}.\textit{e}.}}

\newcommand {\wrt}       {{\textrm{w}.\textrm{r}.\textrm{t}.}}

\newcommand {\at}   {\,}               
\newcommand {\of}   {{:}}              
\newcommand {\Type} {{\mathsf {type}}} 
\newcommand {\sig}  {{\mathsf {sig}}}  

\newcommand {\Dom} {{\mathsf {Dom}}} 
\newcommand {\FV}  {{\mathsf {Fv}}}  

\newcommand {\App}    [2] {{#1} \at {#2}}                         
\newcommand {\Prod}   [3] {\Pi {#1} \of {#2}.{#3}}            
\newcommand {\Abs}    [3] {\lambda {#1} \of {#2}.{#3}}   

\newcommand {\Lock}   [4] {{\mathcal{L}}^{#1}_{#2, #3}[{#4}]} 
\newcommand {\LockC} [3] {{\mathcal{L}}^{#1}_{#2}      [{#3}]} 

\newcommand {\Unlock} [4] {{\mathcal{U}}^{#1}_{#2, #3}[{#4}]} 

\newcommand {\VDASH}  {\vdash}
\newcommand {\VDASHS} {\vdash_\Sigma} 
\newcommand {\VDASHO} {\vdash_\Omega} 
\newcommand {\VDASHEAL} {\vdash_{EAL}} 
\newcommand {\VDASHSEAL} {\vdash_{\Sigma_{{EAL}}}} 
\newcommand {\VDASHFP} {\vdash_{\sf FPST}} 

\newcommand {\SIG} {{\mathcal S}}    
\newcommand {\CTX} {{\mathcal C}}    
\newcommand {\KK}  {{\mathcal K}}    
\newcommand {\FF}  {{\mathcal F}}    
\newcommand {\OO}  {{\mathcal O}}    

\newcommand {\AF}  {{\mathcal{F}_a}} 

\newcommand {\AO}  {{\mathcal{O}_a}} 

\renewcommand {\P} {\mathcal{P}} 

\newcommand   {\U} {\mathcal{U}} 

\newcommand {\eqdef} {\stackrel{\tiny \Delta}{=}}

\newcommand {\ELNF}  {$\beta\eta$-lnf}   
\newcommand {\Subst} {{\mathcal S}}

\newcommand{\rew}[1]  {\hspace{-#1mm}}
\newcommand{\fwd}[1]  {\hspace{#1mm}}
\newcommand{\up}[1]   {\vspace{-#1mm}}
\newcommand{\down}[1] {\vspace{#1mm}}

\newcommand{\ROU}	{(O{\cdot}Unlock)}		

\def \LF	{\mbox {{\sf LF}}}

\title{Gluing together Proof Environments: \\ Canonical extensions of\\
  LF Type Theories featuring \emph{Locks}\footnote{The work
    presented in this paper was partially supported by the Serbian
    Ministry of Education, Science, and Technological Development,
    projects ON174026 and III44006.}}

\author{Furio Honsell
\institute{Department of Mathematics and Computer Science\\
University of Udine, Italy}
\email{furio.honsell@uniud.it}
\and
Luigi Liquori
\institute{Inria Sophia Antipolis M\'editerran\'ee, France}
\email{luigi.liquori@inria.fr}
\and Petar Maksimovi\'{c}
\institute{Inria Rennes Bretagne Atlantique, France}
\institute{Mathematical Institute of the Serbian Academy\\ of Sciences
  and Arts, Serbia}
\email{petar.maksimovic@inria.fr}
\and Ivan Scagnetto
\institute{Department of Mathematics and Computer Science\\
University of Udine, Italy}
\email{ivan.scagnetto@uniud.it}
}
\def\titlerunning{Gluing together Proof Environments: \CLLFP\ \& \CLLFPQ}
\def\authorrunning{F. Honsell, L. Liquori, P. Maksimovi\'{c}, I. Scagnetto}

\begin{document}

\maketitle

\begin{abstract}
  We present two extensions of the \LF\ Constructive Type Theory
  featuring monadic \emph{locks}. A lock is a monadic type construct
  that captures the effect of an \emph{external call to an
    oracle}. Such calls are the basic tool for \emph{gluing together}
  diverse Type Theories and proof development environments.  The
  oracle can be invoked either to check that a constraint holds or to
  provide a suitable witness. The systems are presented in the
  \emph{canonical style} developed by the CMU School. The first
  system, \CLLFP, is the canonical version of the system \LLFP,
  presented earlier by the authors. The second system, \CLLFPQ,
  features the possibility of invoking the oracle to obtain a witness
  satisfying a given constraint. We discuss encodings of Fitch-Prawitz
  Set theory, call-by-value $\lambda$-calculi, and systems of Light
  Linear Logic. Finally, we show how to use Fitch-Prawitz Set Theory
  to define a type system that types precisely the strongly
  normalizing terms.
\end{abstract}


\section{Introduction}\label{sec:introduction}
In recent years, the authors have introduced in a series of papers
\cite{HLLMS12,Honsell:2013:YFP:2503887.2503896,llfp-mfcs2014,HLMS15}
various extensions of the Constructive Type Theory \LF, with the goal
of defining a simple \emph{Universal Meta-language} that can support
the effect of \emph{gluing together}, \ie\ interconnecting, different
type systems and proof development environments.

The basic idea underpinning these logical frameworks is to allow for
the user to express explicitly, in an \LF\ type-theoretic framework
the \emph{invocation}, and uniform \emph{recording} of the
\emph{effect}, of external tools by means of a new \emph{monadic}
type-constructor $\Lock{\P}{M}{\sigma}{\cdot}$, called a \emph{lock}.
More specifically, locks permit to express the fact that, in order to
obtain a term of a given type, it is necessary to \emph{verify},
first, a constraint ${\P}(\Gamma \VDASHS M : \sigma)$, \ie\ to
\emph{produce} suitable \emph{evidence}. No restrictions are enforced
on producing such evidence. It can be supplied by calling an
\emph{external proof search tool} or an \emph{external oracle}, or
exploiting some other epistemic source, such as diagrams, physical
analogies, or explicit computations according to the \emph{Poincar\'e
  Principle} \cite{bar02}. Thus, by using lock constructors, one can
\emph{factor-out} the goal, produce pieces of evidence using different
proof environments and \emph{glue} them back together, using the
\emph{unlock operator}, which \emph{releases} the locked term in the
calling framework. Clearly, the task of checking the validity of
external evidence rests entirely on the external tool. In our
framework we limit ourselves to recording in the proof term by means
of an $\cal U$-destructor this recourse to an external tool.

One of the original contributions of this paper is that we show how
locks can delegate to external tools not only the task of producing
suitable evidence but also that of exhibiting suitable
\emph{witnesses}, to be further used in the calling environment. This
feature is exhibited by \CLLFPQ\ (see Section~\ref{sec:cllfpq}).

Locks subsume different \emph{proof attitudes}, such as
proof-irrelevant approaches, where one is only interested in knowing
that evidence does exist, or approaches relying on powerful
terminating metalanguages. Indeed, locks allow for a straightforward
accommodation of many different \emph{proof cultures} within a single
Logical Framework; which otherwise can be embedded only very
deeply~\cite{deep,hirschkoff:bisimproofs} or axiomatically
\cite{HMS-01}.

Differently from our earlier work, we focus in this paper only on
systems presented in the \emph{canonical format} introduced by the CMU
school \cite{watkins-02,HarperLicata-jfp-07}. This format is
syntax-directed and produces a unique derivation for each derivable
judgement. Terms are all in normal form and equality rules are
replaced by \emph{hereditary substitution}. We present the systems in
canonical form, since this format streamlines the proof of adequacy
theorems.

First, we present the very expressive system \CLLFP\ and discuss the
relationship to its non-canonical counterpart \LLFP\ in \cite{HLMS15},
where we introduced \emph{lock-types} following the paradigm of
Constructive Type Theory (\ala\ Martin-L\"of), via
\emph{introduction}, \emph{elimination}, and \emph{equality rules}.
This paradigm needs to be rephrased for the canonical format used
here. Introduction rules correspond to \emph{type checking} rules of
\emph{canonical objects}, whereas elimination rules correspond to
\emph{type synthesis} rules of \emph{atomic objects}. Equality rules
are rendered via the rules of \emph{hereditary substitution}.  In
particular, we introduce a \emph{lock constructor} for building
canonical objects $\Lock{\P}{N}{\sigma}{M}$ of type
$\Lock{\P}{N}{\sigma}{\rho}$, via the \emph{type checking rule}
$(O{\cdot}Lock)$. Correspondingly, we introduce an \emph{unlock
  destructor}, $\Unlock{\P}{N}{\sigma}{M}$, and an \emph{atomic rule}
$\ROU$, allowing elimination, in the hereditary substitution rules, of
the lock-type constructor, under the condition that a specific
predicate $\P$ is verified, possibly \emph{externally}, on a
judgement:
$$
\infer[(O{\cdot}Lock)]
{\Gamma \VDASHS {\Lock \P {N} {\sigma} {M}} \Leftarrow {\Lock {\P} {N} {\sigma} {\rho}}}
{\Gamma \VDASHS {M} \Leftarrow {\rho} & \Gamma \VDASHS {N}
  \Leftarrow {\sigma}}
\quad
\infer[(O{\cdot}Unlock)]
{\Gamma \VDASHS {\Unlock \P {N} {\sigma} {A}} \Rightarrow {\rho}}
{
    \Gamma \VDASHS {A} \Rightarrow {\Lock \P {N} {\sigma} {\rho}} 
 & \Gamma \VDASHS {N} \Leftarrow {\sigma} &  \P(\Gamma \VDASHS {N} \Leftarrow {\sigma})
}
$$

\noindent Capitalizing on the monadic nature of the lock constructor,
as we did for the systems in \cite{llfp-mfcs2014,HLMS15}, one can
use locked terms without necessarily establishing the
predicate, provided an \emph{outermost} lock is present. 
This increases the expressivity of the system, and allows
for reasoning under the assumption that the verification is
successful, as well as for postponing and reducing the number of verifications.  The rules which make all this work are:
$$
\begin{array}{r}
\infer[(F{\cdot}Nested{\cdot}Unlock)] {\Gamma \VDASHS {\Lock {\P} {S}
    {\sigma} {\rho'}\ \Type}}
 {
                                 \Gamma, x\of\tau \VDASHS {\Lock \P {S}
     {\sigma} {\rho}}\  \Type & 
                                 \Gamma
     \VDASHS{A}\Rightarrow{\Lock{\P}{S}{\sigma}{\tau}}& 
                                 \rho[ \Unlock {\P} {S}
                                 {\sigma}{A}/x]_{(\tau)^-}^F= \rho'
}
\\[3mm]
\infer[(O{\cdot}Nested{\cdot}Unlock)] {\Gamma \VDASHS {\Lock
    {\P} {S} {\sigma} {M'}}\Leftarrow {\Lock {\P} {S} {\sigma}
    {\rho'}}} {\begin{array}{l@{\qquad}l} \Gamma, x\of\tau \VDASHS
                 \Lock \P {S} {\sigma} {M}\Leftarrow \Lock \P {S} {\sigma} {\rho} &
                 \Gamma \VDASHS {A} \Rightarrow {\Lock \P {S} {\sigma} {\tau}}\\[1mm]
                 \rho[ \Unlock {\P} {S} {\sigma}{A}/x]_{(\tau)^-}^F =
                 \rho' & M[ \Unlock {\P} {S}
                         {\sigma}{A}/x]_{(\tau)^-}^O = M'
               \end{array}}
\end{array}
$$
\noindent The (\emph{O$\cdot$Nested$\cdot$Unlock})-rule is the counterpart of
the elimination rule for monads, once we realize that the standard destructor of monads (see, e.g., \cite{Moggi-Computationallambda})
$let_{T_{{\mathcal P}(\Gamma \VDASH S : \sigma )}} x=A\ in \ N$ can be
replaced, in our context, by $ N[\Unlock{\P}{S}{\sigma}{A}/ x]$. And this holds since
the $\Lock{\P}{S}{\sigma}{\cdot}$-monad satisfies the property
$let_{T_\P}\ x=M\ in \ N\rightarrow N $ if $ x \notin \FV(N)$,
provided $x$ occurs \emph{guarded} in $N$, \ie\ within subterms of the
appropriate lock-type.  The rule
(\emph{F$\cdot$Nested$\cdot$Unlock}) takes care of elimination at the
level of types.

We proceed then to introduce \CLLFPQ. Syntactically, it might appear
as a minor variation of \CLLFP, but the lock constructor is used here
to express the \emph{request} for a witness satisfying a given
property, which is then \emph{replaced} by the unlock operation. In
\CLLFPQ, the lock acts as a \emph{binding operator} and the unlock as an
\emph{application}.

To illustrate the expressive power of \CLLFP\ and \CLLFPQ\ we discuss
various challenging encodings of subtle logical systems, as well as
some novel applications. First, we encode in \CLLFP\ Fitch-Prawitz
consistent Set-Theory (\FPST), as presented in \cite{prawitz}, and to
illustrate its expressive power, we show, by way of example, how
it can type all strongly normalizing terms. Next, we give signatures
in \CLLFP\ of a strongly normalizing $\lambda$-calculus and a system of
Light Linear Logic \cite{DBLP:conf/lics/BaillotCL07}. Finally, in
Section~\ref{subsec:square_roots}, we show how to encode functions in
\CLLFPQ.

The paper is organized as follows: in Section~\ref{sec:canonical} we
present the syntax, the type system and the metatheory of \CLLFP,
whereas \CLLFPQ\ is introduced in
Section~\ref{sec:cllfpq}. Section~\ref{sec:case_studies} is devoted to
the presentation and discussion of case studies. Finally, connections
with related work in the literature appear in
Section~\ref{sec:relwork}.

\section{The Canonical System \CLLFP} \label{sec:canonical}
In this section, we discuss the \emph{canonical} counterpart of \LLFP
\cite{HLMS15}, \ie\ \CLLFP, in the style of
\cite{watkins-02,HarperLicata-jfp-07}. This approach amounts to
restricting the language only to terms in long $\beta\eta$-normal
form. These are the normal forms of the original system which are
normal also \wrt\ typed $\eta$-like expansion rules, namely
$M \rightarrow \lambda x\of\sigma.Mx$ and
$M \rightarrow \Lock{\P}{N}{\sigma}{\Unlock{\P}{N}{\sigma}{M}}$ if $M$ is atomic. The added value of
canonical systems such as \CLLFP\ is that one can streamline results
of adequacy for encoded systems. Indeed, reductions in the
                        meta-language of non-canonical terms reflect only the history of how
the proof was developed using lemmata.

\begin{figure}[t!]
 \up{2}
 {\small
 \begin{center}
   $
   \begin{array}{rcl@{\hspace{0.9cm}}rcl@{\hspace{0.9cm}}l}

     K & \in & \KK & K & ::=
     & \Type \mid {\Prod x \sigma K} & \mbox{\em Kinds}
                                            \\[1mm]

     \alpha & \in & \AF & \alpha & ::=
     & a \mid {\App \alpha N} & \mbox{\em Atomic Families}
     \\[1mm]

     \sigma, \tau, \rho & \in & \FF & \sigma & ::=
     & \alpha \mid {\Prod x {\sigma} {\tau}} \mid {\Lock {\P} N
       {\sigma} {\rho}}  & \mbox{\em Canonical Families}
     \\[1mm]

     A & \in & \AO & A & ::=
     & c \mid x \mid {\App {A} {M}} \mid {\Unlock {\P} {N} \sigma
       {A}} &  \mbox{\em Atomic Objects}
     \\[1mm]

     M, N & \in & \OO & M & ::=
     & A \mid {\Abs x \sigma M} \mid {\Lock {\P} {N} {\sigma} {M}} &
                                                                     \mbox{\em Canonical Objects}
     \\[1mm]
     \Sigma & \in & \SIG & \Sigma & ::= &
                                          \emptyset \mid \Sigma,a \of K \mid \Sigma, c \of \sigma &
                                                                                                    \mbox{\em
                                                                                                    Signatures}

     \\[1mm]
     %
     \Gamma & \in & \CTX & \Gamma & ::= &
                                          \emptyset \mid \Gamma, x\of \sigma &
                                                                               \mbox{\em
                                                                               Contexts}

 \end{array}
 $
 \end{center} }
 \up{4}
 \caption{Syntax of \CLLFP} \up{5}
 \label{fig:cllfsyntax}
\end{figure}

\begin{figure}[t!]
{\small
  \up{2}
  \noindent {\sf ~Valid signatures}
  \\
  $
  \infer[(S{\cdot}Empty)]
  {\emptyset\ \sig}
  {}
  \hfill
  %
  \infer[(S{\cdot}Kind)]
  {\Sigma, a \of K\ \sig}
  {
     \Sigma\ \sig & 
     \VDASHS K & a \not \in \Dom(\Sigma)
   }
 \hfill
 %
 \infer[(S{\cdot}T\!ype)]
 {\Sigma, c \of \sigma\ \sig}
 {
    \Sigma\ \sig & 
    \VDASHS \sigma\ \Type &
    c \not \in \Dom(\Sigma)
  }
$
\begin{center}
  $
  \begin{array}[t]{l}
    \mbox{\sf Kind~rules}\hfill
    \\[1mm]

    \infer[(K{\cdot}T\!ype)]
    {\Gamma \VDASHS \Type}
    {\VDASHS \Gamma}
    \\[2mm]

    \infer[(K{\cdot}Pi)]
    {\Gamma \VDASHS {\Prod x \sigma K}}
    {\Gamma, x \of \sigma \VDASHS K}
    \\[4mm]

     \mbox{\sf Atomic Family~rules}\hfill
     \\[2mm]

     \infer[(A{\cdot}Const)]
     {\Gamma \VDASHS a \Rightarrow K}
     {\VDASHS \Gamma & a \of K \in \Sigma}
     \\[2mm]

     \infer[(A{\cdot}App)]
     {\Gamma \VDASHS \alpha \at M \Rightarrow K}
     {\begin{array}{l}
         \Gamma \VDASHS {\alpha} \Rightarrow {\Prod x {\sigma} {K_1}} \\
         \Gamma \VDASHS M \Leftarrow {\sigma} \\
         K_1[M/x]^K_{(\sigma)^-}=K
       \end{array}}
     \\[6mm]

     \mbox{\sf Canonical Family~rules}\hfill
     \\[2mm]

     \infer[(F{\cdot}Atom)]
     {\Gamma \VDASHS \alpha\ \Type}
     {\Gamma\VDASHS \alpha \Rightarrow \Type}
     \\[2mm]

     \infer[(F{\cdot}Pi)]
     {\Gamma \VDASHS {\Prod x {\sigma} {\tau}}\ \Type}
     {\Gamma, x \of {\sigma} \VDASHS {\tau}\ \Type}
     \\[2mm]

     \infer[(F{\cdot}Lock)]
     {\Gamma \VDASHS {\Lock {\P} N {\sigma} {\rho}}\ \Type}
     {\Gamma \VDASHS \rho\ \Type & \Gamma \VDASHS N \Leftarrow {\sigma}}
     \\[2mm]

     \infer[(F{\cdot}Nested{\cdot}Unlock)]
     {\Gamma \VDASHS {\Lock {\P} {S} {\sigma} {\rho'}\  \Type}}
     {\begin{array}{l}
         \Gamma, x: \tau \VDASHS {\Lock \P {S} {\sigma} {\rho}}\  \Type\\
         \Gamma \VDASHS{A}\Rightarrow{\Lock{\P}{S}{\sigma}{\tau}}\\
         \rho[ \Unlock {\P} {S} {\sigma}{A}/x]_{(\tau)^-}^F= \rho'
       \end{array}}
   \end{array}
  \hfill 
   \begin{array}[t]{l}
    \mbox{\sf Context~rules}\hfill
    \\[1mm]

    \infer[(C{\cdot}Empty)]
    {\VDASHS \emptyset}
    {\Sigma\ \sig}
    \\[2mm]

    \infer[(C{\cdot}T\!ype)]
    {\VDASHS \Gamma, x \of \sigma}
    {
        \VDASHS \Gamma & \Gamma \VDASHS \sigma\  \Type & 
        x \not \in \Dom(\Gamma)
        }
    \\[4mm]

     \mbox{\sf Atomic Object~rules}\hfill
     \\[2mm]

     \infer[(O{\cdot}Const)]
     {\Gamma \VDASHS c \Rightarrow \sigma}
     {\VDASHS \Gamma & c \of \sigma \in \Sigma}
     \\[2mm]

     \infer[(O{\cdot}Var)]
     {\Gamma \VDASHS x \Rightarrow \sigma}
     {\VDASHS \Gamma & x \of \sigma \in \Gamma}
     \\[2mm]

     \infer[(O{\cdot}App)]
     {\Gamma \VDASHS {A} \at {M} \Rightarrow \tau}
     {\begin{array}{l}
       \Gamma \VDASHS {A} \Rightarrow {\Prod x {\sigma} {\tau_1}} \\
       {\Gamma\VDASHS {M} \Leftarrow {\sigma}} \; \; {\tau_1[{M}/x]^F_{(\sigma)^-}}={\tau}
     \end{array}}
   \\[2mm]

   \infer[(O{\cdot}Unlock)]
   {\Gamma \VDASHS {\Unlock \P {N} {\sigma} {A}} \Rightarrow {\rho}}
   {\begin{array}{l}
       \Gamma \VDASHS {A} \Rightarrow {\Lock \P {N} {\sigma} {\rho}} \\
       \Gamma \VDASHS {N} \Leftarrow {\sigma} \; \; \P(\Gamma \VDASHS {N} \Leftarrow {\sigma})
     \end{array}}
   \\[6mm]

   \mbox{\sf Canonical Object~rules}\hfill
   \\[1mm]

   \infer[(O{\cdot}Atom)]
   {\Gamma \VDASHS A \Leftarrow \alpha}
   {\Gamma\VDASHS A \Rightarrow \alpha}
   \\[2mm]

   \infer[(O{\cdot}Abs)]
   {\Gamma \VDASHS {\Abs x {\sigma} M} \Leftarrow {\Prod x {\sigma} {\tau}}}
   {\Gamma, x \of {\sigma} \VDASHS M \Leftarrow {\tau}}
   \\[2mm]

   \infer[(O{\cdot}Lock)]
   {\Gamma \VDASHS {\Lock \P {N} {\sigma} {M}} \Leftarrow {\Lock {\P} {N} {\sigma} {\rho}}}
   {\Gamma \VDASHS {M} \Leftarrow {\rho} & \Gamma \VDASHS {N}
                                           \Leftarrow {\sigma}}
   \\[2mm]
 \end{array}
 $

 $
   \hfill \infer[(O{\cdot}Nested{\cdot}Unlock)]
   {\Gamma \VDASHS {\Lock {\P} {S} {\sigma} {M'}}\Leftarrow {\Lock {\P} {S} {\sigma} {\rho'}}}
   {\begin{array}{l@{\quad}l}
       \Gamma, x\of\tau \VDASHS \Lock \P {S} {\sigma} {M}\Leftarrow
       \Lock \P {S} {\sigma} {\rho} &
       \Gamma \VDASHS {A} \Rightarrow {\Lock \P {S} {\sigma} {\tau}}\\
     \rho[ \Unlock {\P} {S} {\sigma}{A}/x]_{(\tau)^-}^F = \rho' &
     M[ \Unlock {\P} {S} {\sigma}{A}/x]_{(\tau)^-}^O = M'
   \end{array}}
   $
\end{center}}
\up{4}
\caption{The \CLLFP\ Type System} \up{5}
\label{fig:cllftypesys}
\end{figure}

\subsection{Syntax and Type System for \CLLFP}\label{subsec:cllfp_syntax}
The syntax of \CLLFP\ is presented in Figure~\ref{fig:cllfsyntax}. The
type system for \CLLFP\ is shown in Figure~\ref{fig:cllftypesys}.
The judgements of $\CLLFP$ are the following:

\up{1}
{\small
$$
\begin{array}{rcl@{\hspace{0.9cm}}l}
 & \Sigma & \sig & \mbox{$\Sigma$ is a valid signature} \\
 & \VDASHS & \Gamma & \mbox{$\Gamma$ is a valid context in
  $\Sigma$} \\
  \Gamma & \VDASHS & K & \mbox{$K$ is a kind in $\Gamma$ and
  $\Sigma$} \\
 \Gamma & \VDASHS & \sigma\ \Type
 & \mbox{$\sigma$ is a canonical family in $\Gamma$ and $\Sigma$} \\
 \Gamma & \VDASHS & \alpha \Rightarrow K
 & \mbox{$K$ is the kind of the atomic family  $\alpha$ in $\Gamma$ and $\Sigma$} \\
 \Gamma & \VDASHS & M \Leftarrow \sigma
 & \mbox{$M$ is a canonical term of type $\sigma$ in $\Gamma$ and $\Sigma$} \\
 \Gamma & \VDASHS & A \Rightarrow \sigma
 & \mbox{$\sigma$ is the type of the atomic term $A$ in $\Gamma$ and $\Sigma$}
\end{array}
$$
}
The judgements $\Sigma\ \sig $, and $\VDASHS \Gamma$, and
$\Gamma \VDASHS K$ are as in Section~2.1 of \cite{HLLMSJ12}, whereas
the remaining ones are peculiar to the canonical style.  Informally,
the judgment $\Gamma \VDASHS M \Leftarrow \sigma$ uses $\sigma$ to
check the type of the canonical term $M$, while the judgment
$\Gamma\VDASHS A \Rightarrow \sigma$ uses the type information
contained in the atomic term $A$ and $\Gamma$ to synthesize $\sigma$.
Predicates $\P$ in \CLLFP\ are defined on judgements of the shape
$\Gamma \VDASHS M\Leftarrow \sigma$.

There are two rules whose conclusion is the lock constructor
$\Lock{\P}{S}{\sigma}{\cdot}$. But nevertheless, this system is still
\emph{syntax directed}: when there are subterms of the form
$\Unlock{\P}{S}{\sigma}{A}$ in either $M'$ or $\rho'$, the type
checking algorithm always tries to apply the
$(O\cdot Nested\cdot Unlock)$ rule. If this is not possible, it
applies instead the $(O\cdot Lock)$ rule.

The type system makes use, in the rules $({A{\cdot}App})$ and
$(F{\cdot}App)$, of the notion of \emph{Hereditary Substitution},
which computes the normal form resulting from the substitution of one
normal form into another.  The general form of the hereditary
substitution judgement is $T[M/x]^{t}_{\rho}=T'$, where $M$ is the
term being substituted, $x$ is the variable being substituted for, $T$
is the term being substituted into, $T'$ is the result of the
substitution, $\rho$ is the \emph{simple-type} of $M$, and $t$ denotes
the syntactic class (\eg\ atomic families/object, canonical
families/objects, etc.) under consideration. We give the rules of the
Hereditary Substitution in the style of \cite{HarperLicata-jfp-07},
where the erasure function to simple types is necessary to simplify
the proof of termination, which we omit.

The simple-type $\rho$ of $M$ is obtained via the erasure function of
\cite{HarperLicata-jfp-07} (Figure~\ref{fig:erasure}), mapping
dependent into simple-types. The rules for Hereditary Substitution are
presented in Figures~\ref{fig:hsubstapp1} and \ref{fig:hsubstapp2},
using Barendregt's hygiene condition.

Notice that, in the rule $(O{\cdot}Atom)$ of the type system
(Figure~\ref{fig:cllftypesys}), the syntactic restriction of the
classifier to $\alpha$ atomic ensures that canonical forms are
\emph{long $\beta\eta$-normal forms} for the suitable notion of long
$\beta\eta$-normal form, which extends the standard one for
lock-types. For one, the judgement
$x\of \Prod{z}{a}{a} \VDASHS x \Leftarrow \Prod{z}{a}{a}$ is not
derivable, as $ \Prod{z}{a}{a}$ is not atomic, hence
$\VDASHS \lambda x\of (\Prod{z}{a}{a}). x \Leftarrow
\Prod{x}{(\Prod{z}{a}{a})}{\Prod{z}{a}{a}}$
is not derivable. On the other hand,
$\VDASHS {\Abs{x}{(\Prod{z}{a}{a})}{{\Abs{y}{a}{xy}}}} \Leftarrow
\Prod{x}{(\Prod{z}{a}{a})}{\, \Prod{z}{a}{a}}$,
where $a$ is a family constant of kind $\mathit{Type}$, is
derivable. Analogously, for lock-types, the judgement
$x\of \Lock {\P} N {\sigma} {\rho} \VDASHS x \Leftarrow \Lock {\P} N
{\sigma} {\rho}$
is not derivable, since $\Lock {\P} N {\sigma} {\rho}$ is not atomic.
As a consequence, we have that
$\VDASHS \lambda x\of\Lock {\P} N {\sigma} {\rho}. x \Leftarrow \Pi
x\of \Lock {\P} N {\sigma} {\rho}. \Lock {\P} N {\sigma} {\rho}$
is not derivable. However,
$x\of \Lock {\P} N {\sigma} {\rho} \VDASHS \Lock {\P} N {\sigma} {
  \Unlock {\P} N {\sigma} {x} } \Leftarrow \Lock {\P} N {\sigma}
{\rho}$
is derivable, if $\rho$ is atomic. Hence, the judgment
$\VDASHS \lambda x\of \Lock {\P} N {\sigma} {\rho} . \Lock {\P} N
{\sigma} { \Unlock {\P} N {\sigma} {x} } \Leftarrow \Pi x \of \Lock
{\P} N {\sigma} {\rho}. \Lock {\P} N {\sigma} {\rho}$
is derivable.  Note that the unlock constructor takes an \emph{atomic}
term as its main argument, thus avoiding the creation of possible
${\cal L}$-redexes under substitution. Moreover, since unlocks can
only receive locked terms in their body, no abstractions can ever
arise.  In Definition~\ref{def:etaexp}, we formalize the notion of
$\eta$-expansion of a judgement, together with correspondence theorems
between \LLFP\ and \CLLFP.

\begin{figure}[t!]
{\small
 \up{2}
   $
     \infer[]
     {(a)^-=a}
     {}
     \hfill
     \infer[]
     {(\alpha \at M)^-=\rho}
     {(\alpha)^-=\rho}
     \hfill
     \infer[]
     {({\Prod x {\sigma} \tau})^-=\rho_1 \to\rho_2}
     {(\sigma)^-=\rho_1 & (\tau)^-=\rho_2}
     \hfill
     \infer[]
     {({\Lock {\P} N {\sigma} {\tau}})^-= \Lock {\P} N {\sigma} {\rho}}
     {(\tau)^-=\rho}
   $}
   \up{1}
   \caption{Erasure to simple-types} \up{1}
   \label{fig:erasure}
\end{figure}

\begin{figure}[t!]
{\small
 \begin{center}
   $
   \begin{array}[t]{l}
     \mbox{\sf Substitution in Kinds}\hfill
     \\[2mm]

     \infer[(\Subst{\cdot}K{\cdot}Type)]
     {\Type[M_0/x_0]^K_{\rho_0}=\Type}
     {}

     \hspace{0.3cm}

     \infer[(\Subst{\cdot}K{\cdot}Pi)]
     {({\Prod x \sigma K})[M_0/x_0]^K_{\rho_0}={\Prod x \sigma' K'}}
     {\sigma[M_0/x_0]^F_{\rho_0}=\sigma' & K[M_0/x_0]^K_{\rho_0}=K'}
     \\[2mm]

     \mbox{\sf Substitution in Atomic Families}\hfill
     \\[2mm]

     \infer[(\Subst{\cdot}F{\cdot}Const)]
     {a[M_0/x_0]^f_{\rho_0}=a}
     {}

     \hspace{0.5cm}

     \infer[(\Subst{\cdot}F{\cdot}App)]
     {(\alpha M)[M_0/x_0]^f_{\rho_0}=\alpha'M'}
     {\alpha[M_0/x_0]^f_{\rho_0}=\alpha' & M[M_0/x_0]^O_{\rho_0}=M'}
     \\[2mm]

     \mbox{\sf Substitution in Canonical Families}\hfill
     \\[2mm]

     \infer[(\Subst{\cdot}F{\cdot}Atom)]
     {\alpha[M_0/x_0]^F_{\rho_0}=\alpha'}
     {\alpha[M_0/x_0]^f_{\rho_0}=\alpha'}

     \hspace{0.5cm}

     \infer[(\Subst{\cdot}F{\cdot}Pi)]
     {({\Prod x {\sigma_1} {\sigma_2}})[M_0/x_0]^F_{\rho_0}={\Prod x {\sigma'_1} {\sigma'_2}}}
     {\sigma_1[M_0/x_0]^F_{\rho_0}=\sigma'_1 & \sigma_2[M_0/x_0]^F_{\rho_0}=\sigma'_2}
     \\[2mm]

     \infer[(\Subst{\cdot}F{\cdot}Lock)]
     {{\Lock \P {M_1} {\sigma_1} {\sigma_2}}[M_0/x_0]^F_{\rho_0}={\Lock \P {M'_1} {\sigma'_1} {\sigma'_2}}}
     {\sigma_1[M_0/x_0]^F_{\rho_0}=\sigma'_1 &
       M_1[M_0/x_0]^O_{\rho_0}=M'_1 &
       \sigma_2[M_0/x_0]^F_{\rho_0}=\sigma'_2}
   \end{array}
   $
   \caption{Hereditary substitution, kinds and families of \CLLFP} \up{8}
   \label{fig:hsubstapp1}
 \end{center}}
\end{figure}

We present \CLLFP\ in a fully-typed style, \ie\ \ala\ Church, but we
could also follow~\cite{HarperLicata-jfp-07} and present a version
\ala\ Curry, where the canonical forms $\lambda
x.M$ and ${\LockC {\P} M {N}}$ do not carry type information. The type rules would then
be, \eg:

\down{1}
{\small
\hfill$
     \infer[(O{\cdot}Abs)]
     {\Gamma \VDASHS {\lambda x.M} \Leftarrow {\Prod x {\sigma} {\tau}}}
     {\Gamma, x \of {\sigma} \VDASHS M \Leftarrow {\tau}} \hspace{2cm}
     \infer[(O{\cdot}Lock)]
     {\Gamma \VDASHS {\LockC \P {M} {N}} \Leftarrow {\Lock {\P} {M} {\sigma} {\tau}}}
     {\Gamma  \VDASHS {M} \Leftarrow {\sigma} & \Gamma  \VDASHS {N} \Leftarrow {\tau}}
$\hfill} \down{1}

\noindent This latter syntax is more suitable in implementations
because it simplifies the notation.  Following~\cite{HLLMS12}, we stick to the typeful
syntax because it allows for a more direct comparison with
non-canonical systems. This, however, is technically immaterial. Since
judgements in canonical systems have unique derivations, one can show
by induction on derivations that any provable judgement in the system
where object terms are \ala\ Curry has a \emph{unique} type decoration
of its object subterms, which turns it into a provable judgement in
the version \ala\ Church. Vice versa, any provable judgement in the
version \ala\ Church can forget the types in its object subterms,
yielding a provable judgement in the version \ala\ Curry.

\begin{figure}[t!]
{\small
 \up{2}
 \begin{center}
   $
   \begin{array}[t]{l}

     \mbox{\sf Substitution in Atomic Objects}\hfill
     \\[2mm]

     \infer[\rew{1}(\Subst{\cdot}O{\cdot}Const)]
     {c[M_0/x_0]^o_{\rho_0}=c}
     {}
     \quad 

     \infer[\rew{1} (\Subst{\cdot}O{\cdot}Var{\cdot}H)]
     {x_0[M_0/x_0]^o_{\rho_0}=M_0 : \rho_0}
     {}
     \quad 

     \infer[\rew{1} (\Subst{\cdot}O{\cdot}Var)]
     {x[M_0/x_0]^o_{\rho_0}=x}
     {x\neq x_0}
     \\[2mm]

     \infer[(\Subst{\cdot}O{\cdot}App{\cdot}H)]
     {(A_1 M_2)[M_0/x_0]^o_{\rho_0}=M' : \rho}
     {A_1[M_0/x_0]^o_{\rho_0}={\Abs x {\rho_2} {M'_1}} : {\rho_2}
       \rightarrow \rho &
       M_2[M_0/x_0]^O_{\rho_0}=M'_2 & M'_1[M'_2/x]^O_{\rho_2}=M'}
     \\[2mm]

     \infer[(\Subst{\cdot}O{\cdot}App)]
     {(A_1 M_2)[M_0/x_0]^o_{\rho_0}=A'_1M'_2}
     {A_1[M_0/x_0]^o_{\rho_0}=A'_1 & M_2[M_0/x_0]^O_{\rho_0}=M'_2}
     \\[2mm]

     \infer[(\Subst{\cdot}O{\cdot}Unlock{\cdot}H)]
     {{\Unlock \P {M} {\sigma} {A}}[M_0/x_0]^o_{\rho_0}=M_1 : \rho}
     {\sigma[M_0/x_0]^F_{\rho_0}=\sigma' & M[M_0/x_0]^O_{\rho_0}=M' &
       A[M_0/x_0]^o_{\rho_0}= {\Lock \P {M'} {\sigma'} {M_1}} : {\Lock \P {M'} {\sigma'} {\rho}}    }
     \\[2mm]

     \infer[(\Subst{\cdot}O{\cdot}Unlock)]
     {{\Unlock \P {M} {\sigma} {A}}[M_0/x_0]^o_{\rho_0}={\Unlock \P {M'} {\sigma'} {A'}}}
     {\sigma[M_0/x_0]^F_{\rho_0}=\sigma' & M[M_0/x_0]^O_{\rho_0}=M' &
       A[M_0/x_0]^o_{\rho_0}=A'}
     \\[2mm]

     \mbox{\sf Substitution in Canonical Objects}\hfill
     \\[2mm]

     \infer[\rew{1} (\Subst{\cdot}O{\cdot}R)]
     {A[M_0/x_0]^O_{\rho_0}=A'}
     {A[M_0/x_0]^o_{\rho_0}=A'}

     \quad

     \infer[\rew{1} (\Subst{\cdot}O{\cdot}R{\cdot}H)]
     {A[M_0/x_0]^O_{\rho_0}=M'}
     {A[M_0/x_0]^o_{\rho_0}=M' : \rho}
     \quad 

     \infer[\rew{1} (\Subst{\cdot}O{\cdot}Abs)]
     {{\Abs x {\sigma} M}[M_0/x_0]^O_{\rho_0}= {\Abs x {\sigma} M'}}
     {M[M_0/x_0]^O_{\rho_0}=M'}
     \\[2mm]

     \infer[(\Subst{\cdot}O{\cdot}Lock)]
     {{\Lock \P {M_1} {\sigma_1} {M_2}}[M_0/x_0]^O_{\rho_0}={\Lock \P {M'_1} {\sigma'_1} {M'_2}}}
     {\sigma_1[M_0/x_0]^F_{\rho_0}=\sigma'_1 &
       M_1[M_0/x_0]^O_{\rho_0}=M'_1 &
       M_2[M_0/x_0]^O_{\rho_0}=M'_2}
     \\[2mm]

     \mbox{\sf Substitution in Contexts}\hfill
     \\[2mm]

     \infer[\rew{1}(\Subst{\cdot}Ctxt{\cdot}Empty)]
     {[M_0/x_0]^C_{\rho_0}=\emptyset}
     {}
     \quad  

     \infer[\rew{1}(\Subst{\cdot}Ctxt{\cdot}Term)]
     {(\Gamma, x\of \sigma)[M_0/x_0]^C_{\rho_0}=\Gamma', x\of \sigma'}
     {x_0\neq x & x\not\in \FV(M_0) &
       \Gamma[M_0/x_0]^C_{\rho_0}=\Gamma' &
       \sigma[M_0/x_0]^F_{\rho_0}=\sigma'}
   \end{array}
   $
   \up{1}
   \caption{Hereditary substitution, objects and contexts of \CLLFP} \up{5}
   \label{fig:hsubstapp2}
 \end{center}} \up{3}
\end{figure}

\up{2}
\subsection{The Metatheory of \CLLFP}

For lack of space we omit proofs, but these follow the standard
patterns in \cite{HarperLicata-jfp-07,HLLMSJ12}.  We start by studying
the basic properties of hereditary substitution and the type
system. First of all, we need to assume that the predicates are
\emph{well-behaved} in the sense of \cite{HLLMSJ12}. In the context of
canonical systems, this notion needs to be rephrased as follows:

\begin{definition}[Well-behaved predicates for canonical systems]
  \label{def:wbred}
  A finite set of predicates $\{ \P_i\}_{i\in I}$ is
  \emph{well-behaved} if each $\P$ in the set satisfies the following
  conditions:

  \begin{enumerate}
    \setlength\itemsep{-0.3ex}
  \item {\bf \emph{Closure under signature and context weakening and
        permutation:}}\vspace{-0.5ex}
    \begin{enumerate}
      \setlength\itemsep{-0.3ex}
    \item If $\Sigma$ and $\Omega$ are valid signatures such that
      $\Sigma \subseteq \Omega$ and
      $\P(\Gamma \VDASHS N\Leftarrow\sigma)$, 
      then $\P(\Gamma \VDASHO N\Leftarrow\sigma)$. 
    \item If $\Gamma$ and $\Delta$ are valid contexts such that
      $\Gamma\subseteq \Delta$ and
      $\P(\Gamma \VDASHS N\Leftarrow\sigma)$, 
      then \mbox{$\P(\Delta \VDASHS
        N\Leftarrow\sigma)$.} 
    \end{enumerate}
  \item{\bf \emph{Closure under {hereditary} substitution:}} If
    $\P(\Gamma, x \of \sigma', \Gamma' \VDASHS N {\Leftarrow}\
    \sigma)$ and $\Gamma \VDASHS N' : \sigma'$, 
    then\\
    $\P(\Gamma, \Gamma'[N'/x]{^C_{(\sigma')^-}} \VDASHS N
    [N'/x]^O_{(\sigma')^-} {\Leftarrow}\
    \sigma[N'/x]^F_{(\sigma')^-})$.
  \end{enumerate}
\end{definition}

As canonical systems do not feature reduction, the ``classical'' third
constraint for well-behaved predicates (closure under reduction) is
not needed here. Moreover, the second condition (\emph{closure under
  substitution}) becomes ``closure under hereditary substitution''.
\begin{lemma}[Decidability of hereditary substitution]\hfill
  \begin{enumerate}
  \setlength\itemsep{-0.3ex}
\item For any $T$ in $\{\KK, {\mathcal A}, \FF, \OO, {\mathcal C}\}$,
  and any $M$, $x$, and $\rho$, it is decidable whether there exists a
  $T'$ such that $T[M/x]^m_{\rho}= T'$ or there is no such $T'$.
  \item For any $M$, $x$, $\rho$, and $A$, it is decidable whether
    there exists an $A'$, such that $A[M/x]^o_{\rho}= A'$, or there exist
    $M'$ and $\rho'$, such that $A[M/x]^o_{\rho}= M' : \rho'$, or
    there are no such $A'$ and $M'$.
  \end{enumerate}
\end{lemma}
\begin{lemma}[Head substitution size] If $A[M_0/x_0]_{\rho_0}^{o}=M
  \of \rho$, then $\rho$ is a subexpression of $\rho_0$.\vspace{-1ex}
\end{lemma}
\begin{lemma}[Uniqueness of substitution and synthesis] \hfill \vspace{-0.5ex}
  \begin{enumerate}
  \setlength\itemsep{-0.3ex}
  \item It is not possible that $A[M_0/x_0]_{\rho_0}^{o}=A'$ and
    $A[M_0/x_0]_{\rho_0}^{o}=M \of \rho$.
  \item For any $T$, if $T[M_0/x_0]_{\rho_0}^{m} = T'$, and
    $T[M_0/x_0]_{\rho_0}^{m} = T''$, then $T' = T''$.
  \item If $\Gamma \VDASHS \alpha \Rightarrow K$, and
    $\Gamma \VDASHS \alpha \Rightarrow K'$, then $K = K'$.
  \item If $\Gamma \VDASHS A \Rightarrow \sigma$, and
    $\Gamma \VDASHS A \Rightarrow \sigma'$, then $\sigma = \sigma'$.
  \end{enumerate}\vspace{-1ex}
\end{lemma}
\begin{lemma}[Composition of hereditary substitution]
  Let $x\neq x_0$ and $x\not\in \FV(M_0)$. Then:\vspace{-0.5ex}

  \begin{enumerate}
  \setlength\itemsep{-0.3ex}
\item For all $T_1'$ in $\{\KK,\AF,\FF,\AO,\OO\}$, if
  $M_2[M_0/x_0]^O_{\rho_0}=M_2'$, $T_1[M_2/x]^m_{\rho_2} = T_1'$, and
  $T_1[M_0/x_0]^m_{\rho_0} = T_1''$, then there exists a $T$:
  $T_1'[M_0/x_0]^m_{\rho_0} = T$, and $T_1''[M_2'/x]^m_{\rho_2}=T$.
  \item If $M_2[M_0/x_0]^O_{\rho_0}=M_2'$,
    $A_1[M_2/x]^o_{\rho_2}=M : \rho$, and $A_1[M_0/x_0]^o_{\rho_0}=A$,
    then there exists an $M'$: $M[M_0/x_0]^O_{\rho_0}=M'$,
    and $A[M_2'/x]^o_{\rho_2}=M' : \rho$.
  \item If $M_2[M_0/x_0]^O_{\rho_0}=M_2'$, $A_1[M_2/x]^o_{\rho_2}=A$,
    and $A_1[M_0/x_0]^o_{\rho_0}=M : \rho$, then there exists an $M'$:
    $A[M_0/x_0]^o_{\rho_0}=M' : \rho$, and
    $M[M_2'/x]^O_{\rho_2}=M'$.
    \end{enumerate}
\end{lemma}


\begin{theorem}[Transitivity]
  Let $ \Sigma\ \sig $, $ \VDASHS \Gamma, x_0\of \rho_0, \Gamma'$ and
  $\Gamma \VDASHS M_0 \Leftarrow \rho_0$, and assume that all
  predicates are well-behaved. Then,\vspace{-0.5ex}

  \begin{enumerate}
  \setlength\itemsep{-0.3ex}
  \item There exists a $\Gamma''$: $[M_0/x_0]_{\rho_0}^{C}=
    \Gamma''$ and $\VDASHS \Gamma, \Gamma''$.
  \item If $\Gamma, x_0\of \rho_0, \Gamma' \VDASHS K$ then there
    exists a $K'$: $[M_0/x_0]^K_{\rho_0}K=K'$ and $\Gamma,
    \Gamma'' \VDASHS K'$.
  \item If $\Gamma, x_0\of \rho_0, \Gamma' \VDASHS \sigma \ \Type$,
    then there exists a $\sigma'$:
    $[M_0/x_0]^F_{\rho_0}\sigma=\sigma'$ and $\Gamma, \Gamma''
    \VDASHS \sigma'\ \Type$.
  \item If $\Gamma, x_0\of \rho_0, \Gamma' \VDASHS \sigma \ \Type$
    and $\Gamma, x_0\of \rho_0, \Gamma' \VDASHS M \Leftarrow
    \sigma$, then there exist $\sigma'$ and $M'$:
    $[M_0/x_0]^F_{\rho_0}\sigma=\sigma'$ and
    $[M_0/x_0]^O_{\rho_0}M=M'$ and $\Gamma, \Gamma'' \VDASHS M'
    \Leftarrow \sigma'$.
  \end{enumerate}
\end{theorem}
\begin{theorem}[Decidability of typing]
  \label{thm:dectypcllf}
  If predicates in \CLLFP\ are decidable, then all of the judgements
  of the system are decidable.\vspace{-1ex}
\end{theorem}

We can now precisely state the relationship between \CLLFP\ and the
\LLFP\ system of \cite{HLMS15}:

\begin{theorem}[Soundness] For any predicate $\P$ of \CLLFP, we define
  a corresponding predicate in \LLFP\ as follows:
  $\P (\Gamma \VDASHS M : \sigma)$ holds if and only if
  $\Gamma \VDASHS M: \sigma$ is derivable in \LLFP\ and
  $\P(\Gamma \VDASHS M \Leftarrow \sigma)$ holds in \CLLFP. Then, we
  have:\vspace{-0.5ex}
  \begin{enumerate}
  \setlength\itemsep{-0.3ex}
\item If $ \Sigma\ \sig$ is derivable in \CLLFP, then $\Sigma \ \sig$
  is derivable in \LLFP.
  \item If $\VDASHS \Gamma$ is derivable in \CLLFP, then
    $\VDASHS \Gamma$ is derivable in \LLFP.
  \item If $\Gamma \VDASHS K$ is derivable in \CLLFP, then $\Gamma
    \VDASHS K$ is derivable in \LLFP.
  \item If $\Gamma \VDASHS \alpha \Rightarrow K$ is derivable in
    \CLLFP, then $\Gamma \VDASHS \alpha : K$ is derivable in \LLFP.
  \item If $\Gamma \VDASHS \sigma \ \Type$ is derivable in \CLLFP, then
    $\Gamma \VDASHS \sigma : \Type$ is derivable in \LLFP.
  \item If $\Gamma \VDASHS A \Rightarrow \sigma$ is derivable in
    \CLLFP, then $\Gamma \VDASHS A : \sigma$ is derivable in \LLFP.
  \item If $\Gamma \VDASHS M \Leftarrow \sigma$ is derivable in \CLLFP,
    then $\Gamma \VDASHS M : \sigma$ is derivable in \LLFP.
  \end{enumerate}
\end{theorem}

Vice versa, all \LLFP\ judgements in \emph{long $\beta\eta$-normal
  form} (\ELNF) are derivable in \CLLFP.  The definition of a
judgement in \ELNF\ is based on the following extension of the
standard $\eta$-rule to the lock constructor
$\lambda x\of \sigma. Mx \rightarrow_{\eta} M$ and
$\Lock{\P}{N}{\sigma}{\Unlock{\P}{N}{\sigma}{M}} \rightarrow_{\eta} M
$.
\begin{definition}
  An occurrence $\xi$ of a constant or a variable in a term of an
  \LLFP\ judgement is \emph{fully applied and unlocked} \wrt\ its type
  or kind
  $\Prod{\vv{x}_1}{\vv{\sigma}_1}{\vv{\cal L}_1[\ldots
    \Prod{\vv{x}_n}{\vv{\sigma}_n}{\vv{\cal L}_n [\alpha]} \ldots ]
  }$,
  where $\vv{\cal L}_1, \ldots , \vv{\cal L}_n$ are vectors of locks,
  if $\xi$ appears only in contexts that are of the form
  $\vv{\cal U}_n [ (\ldots (\vv{\cal U}_1[\xi \vv{M}_1])\ldots )
  \vv{M}_n]$, where
  $\vv{M}_1, \ldots, \vv{M}_n$,
  $ \vv{\cal U}_1, \ldots , \vv{\cal U}_n$ have the same arities of
  the corresponding vectors of $\Pi$'s and locks.
\end{definition}
\begin{definition}[Judgements in long $\beta\eta$-normal form]
  \hfill \label{def:etaexp}\vspace{-0.5ex}
  \begin{enumerate}
  \setlength\itemsep{-0.3ex}
  \item A term $T$ in a judgement is in \emph{\ELNF} if $T$ is in
    normal form and every constant and variable occurrence in $T$ is
    fully applied and unlocked \wrt\ its classifier in the judgement.
  \item A judgement is in \emph{\ELNF} if all terms appearing in it are
    in \ELNF.
  \end{enumerate}
\end{definition}

\begin{theorem}[Correspondence]\label{thm:correspond} Assume that all
  predicates in \LLFP\ are well-behaved, according to Definition 2.1
  \cite{HLLMSJ12}. For any predicate $\P$ in \LLFP, we define a
  corresponding predicate in \CLLFP\ with:
  $\P (\Gamma \VDASHS M\Leftarrow \sigma)$ holds if
  $\Gamma \VDASHS M\Leftarrow \sigma$ is derivable in \CLLFP\ and
  $\P (\Gamma \VDASHS M : \sigma)$ holds in \LLFP. Then, we
  have:\vspace{-0.5ex}

  \begin{enumerate}
  \setlength\itemsep{-0.3ex}
\item If $ \Sigma\ \sig$ is in \ELNF\ and is \LLFP-derivable, then
  $\Sigma \ \sig$ is \CLLFP-derivable.
\item If $\VDASHS \Gamma$ is in \ELNF\ and is \LLFP-derivable, then
  $\VDASHS \Gamma$ is \CLLFP-derivable.
\item If $\Gamma \VDASHS K$ is in \ELNF, and is \LLFP-derivable, then
  $\Gamma \VDASHS K$ is \CLLFP-derivable.
\item If $\Gamma \VDASHS \alpha : K$ is in 
  \ELNF\ and is \LLFP-derivable, then
  $\Gamma \VDASHS \alpha \Rightarrow K$ is \CLLFP-derivable.
\item If $\Gamma \VDASHS \sigma \of \Type$ is in \ELNF\ and is
  \LLFP-derivable, then $\Gamma \VDASHS \sigma \ \Type$ is
  \CLLFP-derivable.
\item If $\Gamma \VDASHS A : \alpha$ is in 
  \ELNF\ and is \LLFP-derivable, then
  $\Gamma \VDASHS A \Rightarrow \alpha$ is \CLLFP-derivable.
  \item If $\Gamma \VDASHS M : \sigma$ is in \ELNF\ and is
    \LLFP-derivable, then $\Gamma \VDASHS M \Leftarrow \sigma$ is
    \CLLFP-derivable.
  \end{enumerate}
  %
\end{theorem}
Notice that, by the Correspondence Theorem above, any well-behaved
predicate $\P$ in \LLFP\, in the sense of Definition 2.1
\cite{HLLMSJ12} induces a well-behaved predicate in \CLLFP. Finally,
notice that \emph{not} all \LLFP\ judgements have a corresponding
\ELNF. Namely, the judgement
$x\of \Lock{\P}{N}{\sigma}{\rho} \VDASHS x :
\Lock{\P}{N}{\sigma}{\rho}$
does not admit an $\eta$-expanded normal form when the predicate $\P$
does \emph{not} hold on $N$, as the rule $(O{\cdot}Unlock)$ can be
applied only when the predicate holds.

\up{1.5}
\section{The Type System \CLLFPQ}\label{sec:cllfpq}
The main idea behind \CLLFPQ\ (see Figures~\ref{fig:cllfsyntaxQ},
\ref{fig:cllftypesysQ}, and \ref{fig:hsubstappQ})\footnote{{For lack
    of space, we present in these figures only the categories and
    rules of \CLLFPQ\ that differ from their \CLLFP\ counterparts.}}
is to ``empower'' the framework of \CLLFP\ by \emph{adding} to the
lock/unlock mechanism the possibility to receive from the external
oracle a \emph{witness} satisfying suitable constraints. Thus, we can
pave the way for gluing together different proof development
environments beyond proof irrelevance scenarios. In this context, the
lock constructor behaves as a \emph{binder}.
The new $(O{\cdot}Lock)$ rule is the following:

$$
\infer {\Gamma \VDASHS {\Lock \P {x} {\sigma} {M}} \Leftarrow {\Lock
    {\P} {x} {\sigma} {\rho}}} {\Gamma, x \of \sigma \VDASHS {M}
  \Leftarrow {\rho}}
$$
%
\noindent where the variable $x$ is a placeholder bound in $M$ and
$\rho$, which will be replaced by the concrete term that will be
returned by the external oracle call. The intuitive meaning behind the
$(O{\cdot}Lock)$ rule is, therefore, that of recording the need to
delegate to the external oracle the inference of a suitable witness of
a given type. Indeed, $M$ can be thought of as an ``incomplete'' term
which needs to be completed by an inhabitant of a given type $\sigma$
satisfying the constraint $\P$. The actual term, possibly synthesized
by the external tool, will be ``released'' in \CLLFPQ, by the unlock
constructor in the $(O{\cdot}Unlock)$ rule as follows:
%
$$
\infer {\Gamma \VDASHS {\Unlock \P {N} {\sigma} {A}} \Rightarrow
  {\rho'}} {
  \Gamma \VDASHS {A} \Rightarrow {\Lock \P {x} {\sigma} {\rho}} &
  \rho[N/x]^F_{(\sigma)^-} = \rho' &
              \Gamma \VDASHS {N} \Leftarrow {\sigma} &  \P(\Gamma
              \VDASHS {N} \Leftarrow {\sigma})
}
$$
%
\noindent The term ${\U^\P_{N,\sigma}[M]}$ intuitively means that $N$
is precisely the synthesized term satisfying the constraint
$\P(\Gamma\VDASHS N \Leftarrow \sigma)$ that will replace in \CLLFPQ\
all the free occurrences of $x$ in $\rho$. This replacement is
executed in the (${\cal S}{\cdot}O{\cdot}Unlock{\cdot}H$) hereditary
substitution rule (Figure~\ref{fig:hsubstappQ}).

Similarly to \CLLFP, also in \CLLFPQ\ it is possible to ``postpone''
or delay the verification of an external predicate in a lock, provided
an \emph{outermost} lock is present. Whence, the synthesis of the
actual inhabitant $N$ can be delayed, thanks to the
$(O{\cdot}Nested{\cdot}Unlock)$ rule:
%
$$
  \infer{\Gamma \VDASHS {\Lock {\P} {x} {\sigma} {M'}}\Leftarrow
    {\Lock {\P} {x} {\sigma} {\rho'}}}
  {
     \Gamma, y\of \tau \VDASHS \Lock \P {x} {\sigma} {M}\Leftarrow
     \Lock \P {x} {\sigma} {\rho}
     &\!\!
     \Gamma \VDASHS {A} \Rightarrow {\Lock \P {x} {\sigma}  {\tau}}
     &\!\!
     \rho[ \Unlock {\P} {x} {\sigma}{A}/y]_{(\tau)^-}^{F}  =\rho'
     &\!\!
       M[\Unlock {\P} {x} {\sigma}{A}/y]_{(\tau)^-}^{O} = M'
}
$$
%
\noindent The Metatheory of \CLLFPQ\ follows closely that of \CLLFP\
as far as decidability. We have no correspondence theorem since we did
not introduce a non-canonical variant \CLLFPQ. This could have been
done similarly to \LLFP.

\begin{figure}
{\small
 \up{2}
  \begin{center}
    $
    \begin{array}{rcl@{\hspace{0.9cm}}rcl@{\hspace{0.9cm}}l}
      %
      %


      \sigma, \tau, \rho & \in & \FF & \sigma & ::=
      & \alpha \mid {\Prod x {\sigma} {\tau}} \mid {\Lock {\P} {x} {\sigma} {\rho}} & \mbox{\em Canonical Families}
      \\[1mm]


      M, N & \in & \OO & M & ::=
      & A \mid {\Abs x \sigma M} \mid {\Lock {\P} {x} {\sigma} {M}} & \mbox{\em Canonical Objects}
    \end{array}
    $
  \end{center}} \up{4}
  \caption{\CLLFPQ\ Syntax {--- changes \wrt\ \CLLFP}}
  \label{fig:cllfsyntaxQ}   \up{2}
\end{figure}

\begin{figure}
{\small
  \begin{center}
    $
    \begin{array}[t]{l}

      \mbox{\sf Canonical Family~rules}\hfill
      \\[2mm]

%

      \infer[(F{\cdot}Lock)]
      {\Gamma\VDASHS {\Lock {\P} x {\sigma} {\rho}}\ \Type}
      {\Gamma, x \of {\sigma} \VDASHS \rho\ \Type}
      \\[2mm]

      \infer[(F{\cdot}Nested{\cdot}Unlock)]
      {\Gamma \VDASHS {\Lock {\P} {x} {\sigma} {\rho'} \Type}}
      {\begin{array}{l}
          \Gamma, y: \tau \VDASHS {\Lock \P {x} {\sigma} {\rho}}\  \Type\\
          \Gamma \VDASHS{A}\Rightarrow{\Lock{\P}{x}{\sigma}{\tau}}\\
          \rho[ \Unlock {\P} {x} {\sigma}{A}/y]_{(\tau)^-}^F= \rho'
        \end{array}}
      \\[4mm]
    \end{array}
    \hfill    
    \begin{array}[t]{l}
      \mbox{\sf Atomic Object~rules}\hfill
      \\[2mm]

%


      \infer[(O{\cdot}Unlock)]
      {\Gamma \VDASHS {\Unlock \P {N} {\sigma} {A}} \Rightarrow {\rho'}}
      {\begin{array}{ll}
          \Gamma \VDASHS {A} \Rightarrow {\Lock \P {x} {\sigma}
            {\rho}} & \Gamma \VDASHS {N} \Leftarrow {\sigma} \\
           \P(\Gamma \VDASHS {N} \Leftarrow {\sigma}) & \rho[N/x]^F_{(\sigma)^-} = \rho'
        \end{array}}
      \\[2mm]

      \mbox{\sf Canonical Object~rules}\hfill
      \\[2mm]

%

      \infer[(O{\cdot}Lock)]
      {\Gamma \VDASHS {\Lock \P {x} {\sigma} {M}} \Leftarrow {\Lock {\P} {x} {\sigma} {\rho}}}
      {\Gamma x \of \sigma \VDASHS {M} \Leftarrow {\rho}}
      \\[2mm]
    \end{array}
    $

    $
      \hfill \infer[(O{\cdot}Nested{\cdot}Unlock)]
      {\Gamma \VDASHS {\Lock {\P} {x} {\sigma} {M'}}\Leftarrow {\Lock {\P} {x} {\sigma} {\rho'}}}
      {\begin{array}{l@{\quad}l}\Gamma, y\of\tau \VDASHS \Lock \P {x}
         {\sigma} {M}\Leftarrow \Lock \P {x} {\sigma} {\rho} &
          \Gamma \VDASHS {A} \Rightarrow {\Lock \P {x} {\sigma}  {\tau}}\\[1mm]
          \rho[ \Unlock {\P} {x} {\sigma}{A}/y]_{(\tau)^-}^F = \rho' &
          M[ \Unlock {\P} {x} {\sigma}{A}/y]_{(\tau)^-}^O = M'
        \end{array}}
    $
  \end{center}}
  \up{4}
  \caption{The \CLLFPQ\ Type System {--- changes \wrt\ \CLLFP}}
  \label{fig:cllftypesysQ} \up{2}
\end{figure}

\begin{figure}[t!]
{\small
  \begin{center}
    $
    \begin{array}[t]{l}

      \mbox{\sf Substitution in Canonical Families}\hfill
      \\[2mm]


      {\infer[(\Subst{\cdot}F{\cdot}Lock)]
      {{\Lock \P {x} {\sigma_1} {\sigma_2}}[M_0/x_0]^F_{\rho_0}={\Lock \P {x} {\sigma'_1} {\sigma'_2}}}
      { \sigma_1[M_0/x_0]^F_{\rho_0}=\sigma'_1 & \sigma_2[M_0/x_0]^F_{\rho_0}=\sigma'_2}}\\[2mm]

      \mbox{\sf Substitution in Atomic Objects}\hfill
      \\[2mm]

      \infer[(\Subst{\cdot}O{\cdot}Unlock{\cdot}H)]
      {{\Unlock \P {M} {\sigma} {A}}[M_0/x_0]^o_{\rho_0}=M_2 : \rho}
      {\sigma[M_0/x_0]^F_{\rho_0}=\sigma'& M[M_0/x_0]^o_{\rho_0}= M'&
        M_1[M'/x]^o_{(\sigma')^-} = M_2 &
        A[M_0/x_0]^o_{\rho_0}= {\Lock \P {x} {\sigma'} {M_1}} : {\Lock \P {x} {\sigma'} {\rho}}    }
      \\[2mm]

      \mbox{\sf Substitution in Canonical Objects}\hfill
      \\[2mm]

      \infer[(\Subst{\cdot}O{\cdot}Lock)]
      {{\Lock \P {x} {\sigma_1} {M_1}}[M_0/x_0]^O_{\rho_0}={\Lock \P {x} {\sigma'_1} {{M'_1}}}}
      {\sigma_1[M_0/x_0]^F_{\rho_0}=\sigma'_1 &
        M_1[M_0/x_0]^O_{\rho_0}=M'_1}

    \end{array}
    $
  \end{center}}

\up{5}
\caption{\CLLFPQ\ Hereditary Substitution {--- changes \wrt\ \CLLFP}}
\label{fig:hsubstappQ} \up{4}
\end{figure}

\section{Case studies}\label{sec:case_studies}
In this section, we discuss the encodings of a collection of logical
systems which illustrate the expressive power and the flexibility of
\CLLFP\ and \CLLFPQ. We discuss Fitch-Prawitz Consistent Set theory,
\FPST\ \cite{prawitz}, some applications of \FPST\ to normalizing
$\lambda$-calculus, a system of Light Linear Logic in \CLLFP, and an
the encoding of a \emph{partial} function in \CLLFPQ.

The crucial step in encoding a logical system in \CLLFP\ or \CLLFPQ\
is to define the predicates involved in locks. Predicates defined on
closed terms are usually unproblematic. Difficulties arise in
enforcing the properties of closure under hereditary substitution and
closure under signature and context extension, when predicates are
defined on open terms.  To be able to streamline the definition of
well-behaved predicates we introduce the following:
\begin{definition}
  Given a signature $\Sigma$ let $\Lambda_\Sigma$ (respectively
  $\Lambda_\Sigma^o$) be the set of \LLFP\ terms (respectively
  \emph{closed} \LLFP\ terms) definable using constants from
  $\Sigma$. A term $M$ has a \emph{skeleton} in $\Lambda_\Sigma$ if
  there exists a term $N[x_1, \ldots, x_n]\in \Lambda_\Sigma$, whose
  free variables (called \emph{holes} of the skeleton) are in
  $\{x_1,\ldots, x_n\}$, and there exist terms $M_1,\ldots, M_n$ such
  that $M \equiv N[M_1/x_1, \ldots,M_n/x_n]$.
\end{definition}

\subsection{Fitch Set Theory \ala\ Prawitz - {\sf FPST}} \label{Fitch}
In this section, we present the encoding of a formal system of
remarkable logical as well as historical significance, namely the
system of consistent {\emph{Na\"ive} Set Theory}, \FPST, introduced by
Fitch \cite{fitch}. This system was first presented in Natural
Deduction style by Prawitz \cite{prawitz}. As Na\"ive Set Theory is
inconsistent, to prevent the derivation of inconsistencies from the
unrestricted \emph{abstraction} rule, only normalizable
\emph{deductions} are allowed in \FPST. Of course, this side-condition
is extremely difficult to capture using traditional tools.

In the present context, instead, we can put to use the machinery of
\CLLFP\ to provide an appropriate encoding of \FPST\ where the
\emph{global} normalization constraint is enforced \emph{locally} by
checking the proof-object. This encoding beautifully illustrates the
\emph{bag of tricks} that \CLLFP\ supports. Checking that a proof term
is normalizable would be the obvious predicate to use in the
corresponding lock-type, but this would not be a well-behaved
predicate if free variables, \ie\ assumptions, are not sterilized.  To
this end, 
We introduce a distinction between \emph{generic}
judgements, which cannot be directly utilized in arguments, but which
can be assumed, and \emph{apodictic} judgements, which are directly
involved in proof rules. In order to make use of generic judgements,
one has to downgrade them to an apodictic one. This is achieved by a
suitable coercion function.

\begin{definition}[Fitch Prawitz Set Theory, \FPST]
  For the lack of space, here we only
  give the crucial rules for implication and for
  \emph{set-abstraction} and the corresponding elimination rules of the full system of Fitch (see~\cite{prawitz}), as presented by
  Prawitz:
 $$
  \begin{array}{r@{\fwd{15}}r}
    \rew{5}\infer[\rew{1}(\supset{I})]
    {\Gamma \VDASHFP  A \supset B}
    {\Gamma, A \VDASHFP  B}
    &
    \infer[\rew{1} (\supset{E})]
    {\Gamma \VDASHFP  B}
    {\Gamma \VDASHFP A & \Gamma \VDASHFP  A \supset B  }
    \\[2mm]
    \infer[\rew{1} (\lambda I)]
    { \Gamma \VDASHFP T \in \lambda x. A }
    {\Gamma \VDASHFP A[T/x]}
    &
    \infer[\rew{1} (\lambda E)]
    {\Gamma \VDASHFP A[T/x]}
    { \Gamma \VDASHFP T \in \lambda x. A }
  \end{array}
  $$
\end{definition}
\noindent The intended meaning of the term $\lambda x.A$ is the set
$\{x\ |\ A\}$. In Fitch's system, \FPST, conjunction and universal
quantification are defined as usual, while negation is defined
constructively, but it still allows for the usual definitions of
disjunction and existential quantification.  What makes \FPST\
\emph{consistent} is that not all standard deductions in \FPST\ are
legal. Standard deductions are called \emph{quasi-deductions} in
\FPST. A \emph{legal deduction} in \FPST\ is defined instead, as a
quasi-deduction which is \emph{normalizable} in the standard sense of
Natural Deduction, namely it can be transformed in a derivation where
all elimination rules occur before introductions.

\begin{definition}[\LLFP\ signature $\Sigma_{\sf FPST}$ for Fitch
  Prawitz Set Theory]\label{signature-fitch}
The  following constants are introduced: {
\begin{alltt}
o   : Type \(\fwd{26}\iota\)        : Type
T   : o -> Type \(\fwd{15.5}\delta\)        : \(\Pi\)A:o.  (V(A) -> T(A))
V   : o -> Type \(\fwd{16.5}\mathtt{\lambda_{intro}}\) \,\;: \(\Pi\)A:\(\iota\) ->o.\(\Pi\)x:\(\iota\).T(A\ x) -> T(\(\epsilon\) x (lam A))
lam : (\(\iota\) -> o)-> \(\iota\) \(\fwd{6.5}\mathtt{\lambda_{elim}}\)  : \(\Pi\)A:\(\iota\) ->o.\(\Pi\)x:\(\iota\).T(\(\epsilon\) x (lam A))->T(A\ x)
\(\epsilon\)   : \(\iota\) -> \({\iota}\) -> o \(\fwd{9}\mathtt{\supset_{intro}}\): \(\Pi\)A,B:o.(V(A) -> T(B)) -> (T(A \(\supset\)B))
\(\supset \)  \!: o -> o -> o \(\hfill\mathtt{\supset_{elim}}\) : \(\Pi\)A,B:o.\(\Pi\)x:T(A).\(\Pi\)y:T(A\(\supset\)B) -> \({\Lock{\scriptsize\tt{Fitch}}{\scriptsize{\mathtt{\langle{x},y\rangle}}}{\scriptsize\mathtt{{T(A)}\times\mathtt{T(A{\supset}B)}}}{\mathtt{T(B)}}}\)
\end{alltt}
}
\noindent where {\tt o} is the type of propositions, $\supset$ and the
``membership'' predicate $\epsilon$ are the syntactic constructors for
propositions, {\tt lam} is the ``abstraction'' operator for building
``sets'', $\mathtt{T}$ is the apodictic judgement, $\mathtt{V}$ is the
generic judgement, $\delta$ is the coercion function, and
${\langle} {\tt x,y}{\rangle}$ denotes the encoding of pairs, whose
type is denoted by $\mathtt{\sigma {{\times}} \tau}$, \eg\
$\mathtt{\lambda u\of\sigma \rightarrow \tau \rightarrow \rho.\ u\ x\
  y : (\sigma \rightarrow \tau}$
$\mathtt{\rightarrow \rho)\rightarrow \rho}$.  The predicate in the
lock is defined as follows:

  \down{1}
  {\hfill
    {\tt
      Fitch}($\Gamma\vdash_{\Sigma_{\sf FPST}} \mathtt{{\langle}
      x,y{\rangle}\ \Leftarrow\ T(A) {{\times}} T(A\supset B)}$)
  \hfill} \down{1}

  \noindent it holds iff $\mathtt{x}$ and $\mathtt{y}$ have skeletons
  in $\Lambda_{\Sigma_{\sf FPST}}$, all the holes of which have either type
  $\mathtt{o}$ or are guarded by a $\delta$, and hence have type
  $\mathtt{V(A)}$, and, moreover, the proof derived by combining the
  skeletons of $\mathtt{x}$ and $\mathtt{y}$ is normalizable in the
  natural sense. Clearly, this predicate is only semi-decidable.
\end{definition}
For lack of space, we do not spell out the rules concerning the other
logical operators, because they are all straightforward provided we
use only the apodictic judgement $\mathtt{T(\cdot)}$, but a few
remarks are mandatory. The notion of \emph{normalizable proof} is the
standard notion used in natural deduction. The predicate {\tt Fitch}
is well-behaved because it considers terms only up-to holes in the
skeleton, which can have type {\tt o} or are generic judgements.
Adequacy for this signature can be achieved in the format of \cite{HLLMSJ12}:

\begin{theorem}[Adequacy for Fitch-Prawitz Naive Set Theory]
  If $A_1,\ldots,A_n$ are the atomic formulas occurring in
  $B_1,\ldots,B_m,A$, then $B_1\ldots B_m \vdash_{\sf FPST} A $ iff
  there exists a normalizable $\mathtt{M}$ such that
  $\mathtt{A_1\of o,\ldots,A_n\of o,}$ $\mathtt{x_1\of V(B_1),\ldots,x_m\of
    V(B_m)} \vdash_{\Sigma_{\sf FPST}}\mathtt{M \Leftarrow T(A)}$
  (where $\mathtt{A}$, and $\mathtt{B_i}$ represent the encodings of,
  respectively, $A$ and $B_i$ in \CLLFP, for $1\leq i\leq m$).
\end{theorem}

\subsection{A Type System for strongly normalizing $\lambda$-terms}
Fitch-Prawitz Set Theory, \FPST, is a rather intriguing, albeit
unexplored, set theoretic system. The normalizability criterion for
accepting a quasi-deduction prevents the derivation of contradictions
and hence makes the system consistent. Of course, some intuitive rules
are not derivable. For instance \emph{modus ponens} does not hold and
if $t \in \lambda x. A$ then we do not have necessarily that $A[t/x]$
holds.  Similarly, the \emph{transitivity} property does not
hold. However \FPST\ is a very expressive type system which
``encompasses'' many kinds of quantification, provided normalization
is preserved and Fitch has shown, see \eg\ \cite{fitch}, that a large
portion of ordinary Mathematics can be carried out in \FPST.

In this subsection, we sketch how to use \FPST\ to define a {type
  system} which can type \emph{precisely all} the strongly normalizing
$\lambda$-terms. Namely, we show that in \FPST\ there exists a set
$\Lambda$ to which belong only the strongly normalizing
$\lambda$-terms. We speak of a \emph{type system} because the proof in
\FPST\ that a term belongs to $\Lambda$ is \emph{syntax
  directed}. First we need to be able to define recursive objects in
\FPST. 
We adapt, to \FPST, Prop. 4, Appendix A.1 of
\cite{girard1998light}, originally given by J-Y. Girard for Light
Linear Logic, as: 

\begin{theorem}[Fixpoint]
  Let $A[P,x_1 \ldots,x_n]$ be a formula of \FPST\ with an n-ary
  predicate variable $P$. Then, there exists a formula $B$ of \FPST, such
  that there exists a normalizable deduction in \FPST\ between
  $ A[\lambda x_1\ldots,x_{{n}}. B[x_1,\ldots, x_n],x_1
  \ldots,x_n]$ and $B$, and viceversa.\vspace{-1ex}
\begin{proof}
  Let equality be  Leibniz equality, then, assuming $n=1$,
  define
  $\Lambda \equiv \lambda z. \exists x{.\exists y.}
  z={\langle} x,y{\rangle} \& A[(\lambda
  w. {\langle} w,$ $y{\rangle}\in y),x]$.
  Then ${\langle} x,\Lambda{\rangle} \in \Lambda$ is
  equivalent, in the sense of \FPST, to
  $A[{(}\lambda w. {\langle}
  w,\Lambda{\rangle}\in \Lambda),x]$.
\end{proof}
\end{theorem}

Using the Fixpoint Theorem we define first natural numbers, then a
concrete representation of the terms of $\lambda$-calculus, say
$\Lambda_0$.  Using again the Fixed Point Theorem, we define a
(representation of) the substitution function over terms in
$\Lambda_0$ and finally the set $\Lambda$, such that $ x \in \Lambda$
is equivalent in \FPST\ to
$x \in \Lambda_0 \& \forall y. y \in \Lambda_0 \subset app(x,y) \in
\Lambda$.
Here, $ app(x,y)$ denotes the concrete representation of ``applying''
$x$ to $y$. One can derive in \FPST\ that (a representation of) a
$\lambda$-term, say $M$, belongs to $\Lambda$, only if there is a
normalizable derivation of $M\in \Lambda$. But then it is
straightforward to check that only normalizing terms can be typed in
\FPST\ with $\Lambda$, \ie\ belong to $\Lambda$. There is indeed a
natural reflection of the normalizability of the \FPST\ derivation of
the typing judgement $M\in \Lambda$, and the fact that the term
represented by $M$ is indeed normalizable!

\subsection{A Normalizing call-by-value $\lambda$-calculus}
In this section we sketch how to express in \CLLFP\ a call-by-value
$\lambda$-calculus where $\beta$-reductions fire only if the result is
$normalizing$.

{\begin{definition}[Normalizing call-by-value $\lambda$-calculus,
  $\Sigma_{\lambda N}$]
{
\begin{alltt}

o      : Type       Eq  : o -> o -> Type       app : o -> o -> o
v      : Type       var : v -> o               lam : (v -> o) -> o
\(\mathtt{c_{beta}}\) : \(\Pi\)M:o->o,N:o.\(\Lock{\scriptsize{\P\sp{\tt{N}}}}{{\!\!}\scriptsize\mathtt{\langle{M},N\rangle}}{\scriptsize\mbox{\tt{(o->o)\(\times\)o}}}{\mbox{Eq\ (app\ (lam\
\(\lambda\)x:v.M(var\ x))\ N)\ (M\ N)}}\)
\end{alltt}}

  \noindent where the predicate ${\cal P}^{{\tt N}}$ holds on
  $\Gamma\VDASH_{\Sigma_{\lambda N}}\mathtt{\langle
    M,N\rangle}\Leftarrow$
  \texttt{(o->o)$\times$o} if both $\tt M$ and $\tt N$ have skeletons
  in $\Lambda_{\Sigma_{\lambda N}}$ whose holes are guarded by a {\tt
    var} and, moreover, {\tt M\ N} ``normalizes'', in the intuitive
  sense, outside terms guarded by a {\tt var}.
\end{definition}}
\subsection{Elementary Affine Logic}
In this section we give a {\em shallow} encoding of {\em Elementary
  Affine Logic} as presented in
\cite{DBLP:conf/lics/BaillotCL07}. This example will exemplify how
locks can be used to deal with global syntactic constraints as in the
{\em promotion rule} of Elementary Affine Logic.
\begin{definition}[Elementary Affine Logic
  \cite{DBLP:conf/lics/BaillotCL07}]\label{def:elementary}
  Elementary Affine Logic can be specified by the following rules:

{\small
    \hfill$
    \begin{array}{c}
      \infer[(Var)]
      {A \VDASHEAL A}
      {}
      \qquad
      \infer[(Weak)]
      {\Gamma, A \VDASHEAL B}
      {\Gamma \VDASHEAL B}
      \qquad
      \infer[(Abst)]
      {\Gamma \VDASHEAL  A \multimap B}
      {\Gamma, A \VDASHEAL  B}
      \qquad
      \infer[(Appl)]
      {\Gamma, \Delta \VDASHEAL  B}
      {\Gamma \VDASHEAL A & \Delta \VDASHEAL  A \multimap B  }
      \\[2mm]
      \infer[(Contr)]
      { \Gamma, \Delta \VDASHEAL  B}
      {\Gamma \VDASHEAL  !A &  \Delta, !A, \ldots, !A \VDASHEAL  B }
      \qquad
      \infer[(Prom)]
      { \Gamma_1 \ldots \Gamma_n \VDASHEAL  !A}
      {A_1,\ldots,A_n \VDASHEAL  A &  \Gamma_1 \VDASHEAL  !A_1\   & \ldots &
        \Gamma_n \VDASHEAL  !A_n }
    \end{array}
    $\hfill}

\end{definition}

{\begin{definition}[\LLFP\ signature $\Sigma_{EAL}$ for Elementary
  Affine Logic]\label{signature-elementary}

{
\begin{alltt}

o : Type\hfill T : o -> Type\hfill {V : o -> Type}\hfill \(\multimap \): o -> o -> o \hfill ! : o -> o
\(\mathtt{c_{appl}}\)    \!: \(\Pi\)A,B :o. T(A) -> T(A\( \multimap \)B)-> T(B) \hfill{\(\mathtt{c_{val}}\) \!: \(\Pi\)A:o. V(A) -> T(!A)
\(\mathtt{c_{abstr}}\)   \!: \(\Pi\)A,B :o. \(\Pi\)x:(T(A) -> T(B)) -> \(\Lock{\scriptsize\emph{Light}}{\scriptsize\mbox{\tt{x}}}{\scriptsize\mbox{\tt{T(A)->T(B)}}}{\mbox{\tt{T(A\( \multimap \)B)}}}\)
\(\mathtt{c_{promV_1}}\) : \(\Pi\)A,B :o. \(\Pi\)x:(T(A \(\multimap\) B)) -> \(\Lock{\scriptsize\emph{Closed}}{\scriptsize\mbox{\tt{x}}}{\scriptsize{\mathtt{{T(A\multimap B)}}}}{\mbox{\tt{T(!A) -> V(B)}}}\)
\(\mathtt{c_{promV_2}}\) : \(\Pi\)A,B :o. \(\Pi\)x:(V(A \(\multimap\) B)) -> \(\Lock{\scriptsize\emph{Closed}}{\scriptsize\mbox{\tt{x}}}{\scriptsize{\mathtt{{V(A\multimap B)}}}}{\mbox{\tt{T(!A) -> V(B)}}}\)}
\end{alltt}}

  \noindent where {\tt o} is the type of propositions, $\multimap$ and
  $\mathtt !$ are the obvious syntactic constructors, $\mathtt{T}$ is
  the basic judgement, and $\mathtt{V(\cdot)}$ is an auxiliary
  judgement.  The predicates involved in the locks are defined as
  follows:

  \begin{itemize}
  \setlength\itemsep{-0.3ex}
\item {\tt
    \emph{Light}}($\Gamma\VDASHSEAL \mathtt{x \Leftarrow
    T(A)\rightarrow T(B)}$)
  holds iff if $\mathtt{A}$ is not of the shape $\mathtt{!A}$ then the
  bound variable of $\mathtt{x}$ occurs at most once in the normal
  form of $\mathtt{x}$.
  \item {\tt
      \emph{Closed}}($\Gamma\VDASHSEAL \mathtt{x
      \Leftarrow T(A)}$)
    holds iff the skeleton of {x} contains only free variables of type
    \texttt{o}, \ie no variables of type $\mathtt{T(B)}$, for any
    $\mathtt{B : o}$.
  \end{itemize}
\end{definition}
}%

A few remarks are mandatory. The promotion rule in
\cite{DBLP:conf/lics/BaillotCL07} is in effect a \emph{family} of
natural deduction rules with a growing number of assumptions.  Our
encoding achieves this via the auxiliary judgement $\mathtt{V(\cdot)}$,
the effect of which is self-explanatory.  Adequacy for this signature can be
achieved only in the format of
\cite{HLLMSJ12}, namely:
\begin{theorem}[Adequacy for Elementary Affine Logic]
  if $A_1,\ldots,A_n$ are the atomic formulas occurring in
  $B_1,\ldots,B_m,A$, then $B_1\ldots B_m \VDASHEAL A $ iff there
  exists $\mathtt{M}$ and $\mathtt{A_1\of o,\ldots,A_n\of o,x_1\of}$
  $\mathtt{T(B_1),\ldots,x_m\of T(B_m)} \VDASHSEAL\mathtt{M \Leftarrow
    T(A)}$
  (where $\mathtt{A}$, and $\mathtt{B_i}$ represent the encodings of,
  respectively, $A$ and $B_i$ in \CLLFP, for $1\leq i\leq m$) and all
  variables $\mathtt{x_1 \ldots x_m}$ occurring more than once in
  $\mathtt{M}$ have type of the shape
  $\mathtt{T(B_i)}\equiv \mathtt{T(!C_i)}$ for some suitable formula
  $\mathtt{C}_i$.
\end{theorem}

\noindent The check on the context of the Adequacy Theorem is {\em
  external} to the system \LLFP, but this is in the nature of results
which relate \emph{internal} and \emph{external} concepts. For
example, 
the very concept of \LLFP\ context, which appears in any adequacy
result, is external to \LLFP. Of course, this check is internalized if
the term is closed.

\subsection{Square roots of natural numbers in \CLLFPQ}\label{subsec:square_roots}
It is well-known that logical frameworks based on Constructive Type
Theory do not permit definitions of non-terminating functions (\ie,
all the functions one can encode in such frameworks are total). One
interesting example of \CLLFPQ\ system is the possibility of reasoning
about partial functions by delegating their computation to external
oracles, and getting back their possible outputs, via the lock-unlock
mechanism of \CLLFPQ.

For instance, we can encode natural numbers and compute their square
roots by means of the following signature
($\mathtt{{\langle} x,y{\rangle}}$ denotes the encoding of pairs,
whose type is denoted by $\mathtt{\sigma\times\tau}$, and \texttt{fst}
and \texttt{snd} are the first and second projections, respectively):
{
\begin{alltt}
nat: \(\Type\)\hfill O: nat\hfill S: nat->nat\hfill plus : nat->nat->nat\hfill minus : nat->nat->nat
mult  : nat->nat->nat\hfill   sqroot: nat->nat\hfill        eval : nat->nat->\(\Type\)
sqrt  : \(\Pi\mathtt{x\of{nat}}.\Lock{\textit{SQRT}}{\mathtt{y}}{\mathtt{nat}\times\sigma}{\mathtt{(eval\
(sqroot\ x)\ (fst y))}}\)
\end{alltt}}
  \noindent where \texttt{eval} represents the usual evaluation
  predicate, the variable $\tt y$ is a pair and

   \down{1}
  {\hfill $\sigma \equiv \mathtt{(eval\ (plus\ (minus\ x\ (mult\ z\ z))\ (minus\
    (mult\ z\ z)} \mathtt{\ x)\ O))}$\hfill} \down{1}

\noindent and
${\textit{SQRT}}(\Gamma\VDASHS \mathtt{y} \Leftarrow
\mathtt{nat}\times\sigma)$
holds if and only if the first projection of \texttt{y}
is the minimum number \texttt{N} such that
$\mathtt{(x\dotminus N\ast N)+(N\ast N\dotminus x)=0}$, where $+$ and
* are represented by \texttt{plus} and \texttt{mult}, while
$\mathtt \dotminus$ (represented by \texttt{minus} in our signature)
is defined as follows:

\down{1}
{ \hfill$
\mathtt{x\dotminus y\eqdef } \left\{\begin{array}{l@{\quad}l}
                                         \mathtt{x-y} & \mathtt{if}~\mathtt{x\geq y}\\
                                         \mathtt{0}     & \mathtt{ otherwise}
                                       \end{array}\right.
$\hfill}

\noindent Thus, the specification of \texttt{sqroot} is not explicit
in \CLLFPQ, since it is implicit in the definition of
${\textit{SQRT}}$.

\section{Related work}\label{sec:relwork}
Building a universal framework with the aim of ``gluing'' different
tools and formalisms together is a long standing goal that has been
extensively explored in the inspiring work on Logical Frameworks
by~\cite{BCKL-03,
  pfenning1999system,watkins-02,celf,LF-modulo,dedukti,NPP05:CMTT,Pientka08:DependentBeluga,belugasys,HLL06}. Moreover,
the appealing monadic structure and properties of the lock/unlock
mechanism go back to Moggi's notion of computational monads
\cite{Moggi-Computationallambda}. Indeed, our system can be seen as a
generalization to a family of dependent \emph{lax} operators of
Moggi's \emph{partial} $\lambda$-calculus~\cite{moggi1988partial} and
of the work carried out
in~\cite{fairtlough1997propositional,mendler1991constrained} (which is
also the original source of the term ``lax''). A correspondence
between lax modalities and monads in functional programming was
pointed out in \cite{alechina2001,garg2008indexed}.  On the other
hand, although the connection between constraints and monads in logic
programming was considered in the past, \eg, in
~\cite{NPP05:CMTT,Fairtlough97first-orderlax,fairtlough2001abstraction},
to our knowledge, our systems are the first attempt to establish a
clear correspondence between side conditions and monads in a
\emph{higher-order dependent-type theory} and in logical frameworks.
Of course, there are a lot of interesting points of contact with other
systems in the literature which should be explored. For instance,
in~\cite{NPP05:CMTT}, the authors introduce a contextual modal logic,
where the notion of context is rendered by means of monadic
constructs. We only point out that, as we did in our system,  they
could have also simplified their system by doing away with the {\tt let}
construct in favor of a deeper substitution.
Schr\"oder-Heister has discussed in a number of papers, see \eg\
\cite{schroeder2012proof,schroeder2012honour}, various restrictions
and side conditions on rules and on the nature of assumptions that one
can add to logical systems to prevent the arising of paradoxes. There
are some potential connections between his work and ours. It would be
interesting to compare his requirements on side conditions being
``closed under substitution'' to our notion of \emph{well-behaved}
predicate. Similarly, there are commonalities between his distinction
between \emph{specific} and \emph{unspecific} variables, and our
treatment of free variables in well-behaved predicates.  LFSC,
presented in~\cite{stump-08}, is more reminiscent of our approach as
``it extends \LF\ to allow side conditions to be expressed using a
simple first-order functional programming language''. Indeed, the
author factors the verifications of side-conditions out of the main
proof. The task is delegated to the type checker, which runs the code
associated with the side-condition, verifying that it yields the
expected output. The proposed machinery is focused on providing
improvements for SMT solvers.


{

}

\end{document}